\documentclass[journal]{IEEEtran}
\usepackage{cite}
\usepackage{amsmath,amssymb,amsfonts}
\usepackage{algorithmic}
\usepackage{graphicx}
\usepackage{textcomp}
\usepackage{xcolor}
\usepackage{hyperref}
\usepackage{siunitx}
\usepackage{afterpage}
\usepackage{multirow}
\usepackage{float}
\usepackage[section]{placeins}
\usepackage{placeins}
\usepackage{tabularx}
\usepackage{comment}
\usepackage{booktabs}
\usepackage[ruled,vlined]{algorithm2e}
\SetKwComment{tcc}{}{}
\usepackage{xcolor}
\hypersetup{%
  colorlinks=true,%
  linkcolor={red},
  citecolor={blue},
  urlcolor={blue},
  bookmarksnumbered=true,%
  bookmarksopen=true}
\def\BibTeX{{\rm B\kern-.05em{\sc i\kern-.025em b}\kern-.08em
    T\kern-.1667em\lower.7ex\hbox{E}\kern-.125emX}}

\begin{document}

\title{Deep Learning-Enabled Signal Detection for MIMO-OTFS-Based 6G and Future Wireless Networks}

\author{Emin Akpinar,~Emir Aslandogan,~Burak Ahmet Ozden,~Haci Ilhan,~\IEEEmembership{Senior Member,~IEEE},~Erdogan Aydin
\thanks{E. Akpinar, E. Aslandogan, and H. Ilhan are with Y{\i}ld{\i}z Technical University, Department of Electronics and Communications Engineering, 34220, Davutpasa, Istanbul, Turkey (e-mail: \{emin.akpinar, emira, ilhanh\}@yildiz.edu.tr).}  
\thanks{B. A. Ozden is with the Department of Electrical and Electronics Engineering, Istanbul Medeniyet University, 34857 Istanbul, T\"urkiye, and also with the Department of Computer Engineering, Y{\i}ld{\i}z Technical University, 34220 Istanbul, T\"urkiye (e-mail: bozden@yildiz.edu.tr).}
\thanks{E. Aydin is with the Department of Electrical and Electronics Engineering, Istanbul Medeniyet University, 34857 Istanbul, T\"urkiye (e-mail: erdogan.aydin@medeniyet.edu.tr).}
\thanks{This work was supported by the Scientific and Technological Research Council of Turkey (TUBITAK) under Project 123E513 and Scientific Research Coordination Unit of Y{\i}ld{\i}z Technical University under project FBA-2024-6361.}}

\maketitle

\begin{abstract}
Orthogonal time frequency space (OTFS) modulation stands out as a promising waveform for sixth generation (6G) and beyond wireless communication systems, offering superior performance over conventional methods, particularly in high-mobility scenarios and dispersive channel conditions. Recent research has demonstrated that the reduced computational complexity of deep learning (DL)-based signal detection (SD) methods constitutes a compelling alternative to conventional techniques. In this study, low-complexity DL-based SD methods are proposed for a multiple-input multiple-output (MIMO)-OTFS system and examined under Nakagami-$m$ channel conditions. The symbols obtained from the receiver antennas are combined using maximum ratio combining (MRC) and detected with the help of a DL-based detector implemented with multi-layer perceptron (MLP), convolutional neural network (CNN), and residual network (ResNet). Complexity analysis reveals that the MLP architecture offers significantly lower computational complexity compared to CNN, ResNet, and classical methods such as maximum likelihood detection (MLD). Furthermore, numerical analyses have shown that the proposed DL-based detectors, despite their low complexity, achieve comparable bit error rate (BER) performance to that of a high-performance MLD under various system conditions.
\end{abstract}


\section{Introduction}
\IEEEPARstart{O}{rthogonal} time frequency space (OTFS) modulation is a highly efficient technology for use in communication systems with high-speed users, where a strong Doppler effect is present \cite{11274686, Orthogonal_Time_Frequency_Space_Modulation, Li_Xiang_Yuan_Weijie_2024, New_Generation_of_Modulation_Addressing_the_Challenges, 10794219, 11274790}. Unlike traditional modulation techniques, OTFS applies a series of two-dimensional transformations to a dual dispersive channel. As a result, the channel is transformed into one that exhibits almost no attenuation in the delay-Doppler (DD) domain \cite{Ramachandran_Chockalingam_2018, Hybrid_MAP_and_PIC_Detection_for_OTFS_Modulation}. At high frequencies and user velocities, the transmitter, receiver, and surrounding scatterers show the disruptive effect of Doppler shift depending on their relative motion.  Conventional orthogonal frequency-division multiplexing (OFDM), widely employed in forth-generation (4G), emerging fifth generation (5G) cellular systems, and wireless fidelity (WiFi) networks, experiences performance degradation in such environments due to inter-carrier interference (ICI) \cite{Bit_Error_Rate_OTFS}. This interference is further intensified by the notable disparity in normalized Doppler effects between the highest and lowest subcarriers, thereby complicating the synchronization process \cite{Orthogonal_Time-Frequency_Space_Modulation_A_Promising_Next_Generation_Waveform}. While OFDM uses the time-frequency (TF) domain to modulate data symbols, OTFS uses the DD domain to modulate data symbols in order to eliminate the dispersion effect caused by the channel, which is particularly seen in satellite and vehicle communication systems \cite{Two-Dimensional_Convolutional_Neural_Network-Based_Signal_Detection_for_OTFS_Systems, Data-driven_deep_learning_for_OTFS_detection, IM_OTFS_survey}. 

Although signal detection (SD) is jointly performed in the DD domain, traditional methods such as minimum mean square error (MMSE) \cite{9814638} and maximum a posteriori (MAP) detection \cite{Li_2020} still impose significant computational complexity. The necessity for techniques with reduced complexity has arisen due to this situation \cite{Optimal_Low-Complexity_Orthogonal_Block_Based_Detection_of_OTFS_for_Low-Dispersion_Channels}. Many studies have been conducted regarding OTFS-SD, and these studies addressed various approaches. These involve traditional techniques such as zero-forcing (ZF) \cite{Low-Complexity_Linear_Equalization_for_OTFS_Modulation} and MMSE \cite{8859227}, alongside advanced methodologies like message passing (MP) \cite{8424569, 9369378, 9492800, 9349154, 9508141, 9610105, WU2024948} and its approximations, as well as Bayesian inference techniques \cite{9082873}. Additionally, expectation propagation (EP) \cite{9503406}, Bayesian parallel interference cancellation, and maximal ratio combining (MRC) detection \cite{Low_Complexity_MRC_Detection_for_OTFS_Receiver_With_Oversampling, Low_Complexity_Iterative_Rake_Detector_for_Orthogonal_Time_Frequency_Space_Modulation, OTFS_Based_Receiver_Scheme_with_Multi-Antennas_in_High-Mobility_V2X_Systems, 9293173}, along with various equalisation strategies, have been investigated to improve OTFS-SD performance \cite{9866655}.

Deep learning (DL) based OTFS-SD can be used to design techniques that are more suitable for dynamic channel conditions, thereby reducing computational complexity and improving overall detection efficiency \cite{OTFS_CE_survey}. In \cite{9448630}, a low-complexity OTFS-SD system based on deep neural networks (DNNs) was examined. A symbol-level DNN architecture was proposed to reduce computational complexity, while results were close to full DNN and maximum likelihood detection (MLD) performance. The study further demonstrates that DNNs performing at the symbolic level outperform MLD in channel conditions characterized by non-Gaussian noise. In \cite{9518377}, a DL-based OTFS-SD method utilizing a two-dimensional (2D) convolutional neural network (CNN) and data augmentation with the MP algorithm is proposed, demonstrating improved performance over the MP detector and achieving near-optimal results comparable to the MAP detector with low computational complexity. In \cite{9900413}, a Bayesian parallel interference cancellation network (BPICNet) OTFS detector is proposed that integrates neural networks (NNs), Bayesian inference and parallel interference cancellation. This enhances the performance of detection in environments with rich scattering and demonstrates superiority over state-of-the-art, low-complexity OTFS detectors. In \cite{9814608}, a DL-based OTFS detector integrated with a data augmentation method using a computationally efficient linear detector and a 2D-CNN is proposed to effectively leverage the DD channel and determine the multiple-input multiple-output (MIMO)-OTFS input-output relationship. The study in \cite{Data-driven_deep_learning_for_OTFS_detection} introduces a data-driven DL-based OTFS detection approach. In this approach, ResNet, DenseNet, and residual dense network (RDN) architectures are used to address the challenges posed by channel variations. In \cite{10018844}, a low-complexity OTFS-SD algorithm, ViterbiNet, is proposed, where a NN replaces the log-likelihood computation in the traditional Viterbi algorithm. In addition to maintaining high performance and utilizing the soft-plus activation function to improve training stability with minimal training data, the study shows that the ViterbiNet-based approach can successfully perform OTFS-SD without requiring channel state information (CSI). In \cite{10105487}, an EP-aided model-driven DL framework is proposed for OTFS-SD to mitigate the challenges posed by rich-scattering environments. The study shows that by unfolding each iteration of the EP algorithm into a layer-wise NN with trainable parameters, the proposed approach accelerates convergence and significantly enhances detection performance compared to conventional EP-based methods. In \cite{10154051}, a graph NN (GNN)-assisted detector is proposed for OTFS-SD, where transmit symbols are modeled as graph nodes and processed through aggregation, update, and output modules. According to the study, the suggested GNN-assisted detector outperforms the more advanced detectors by about 1 dB.

In this study, a DL-based low-complexity SD method based on MRC for MIMO-OTFS is proposed. The proposed detection scheme offers a practical alternative to the highly computational MLD method and has been investigated for both single-input single-output (SISO) and MIMO systems, considering different antenna configurations. Furthermore, a comparative analysis is conducted on three DL architectures: MLP, CNN, and ResNet. The analysis indicates that, compared to more complex models such as CNN and ResNet, the MLP architecture demonstrated similar performance in terms of bit error rate (BER) while having lower computational complexity. Additionally, the performance of the proposed approaches is performed on Nakagami-$m$ fading channels to model realistic wireless channel conditions. BER analyses are conducted under different fading conditions, and the robustness of the MIMO-OTFS system against varying channel conditions is confirmed.

The rest of this paper is organized as follows: 
Section \ref{sec:mimo_otfs} presents the MIMO-OTFS system model, detailing the transmitter, Nakagami-m fading channel, and receiver stages. Section \ref{sec:nn_signal_det} describes the proposed deep learning-based signal detection frameworks, including the MLP, CNN, and ResNet architectures used for symbol classification. Section \ref{sec:complexity} provides a comprehensive computational complexity analysis, comparing the real multiplication counts of the proposed DL models against the conventional MLD. Section \ref{sec:num_res_dis} discusses the numerical results, evaluating the BER performance under various antenna configurations and fading conditions. Finally, Section \ref{sec:conclusion} concludes the paper and outlines future research directions.

\textit{Notations:} The key notations used throughout this article are summarized in Table~\ref{tab:notations}.

\begin{table}[t] 
\caption{List of key notations and definitions.}
\label{tab:notations}
\centering
\begin{tabularx}{\columnwidth}{l X}
\toprule
\textbf{Symbol} & \textbf{Description} \\
\midrule
\multicolumn{2}{l}{\textit{General Notation and Operators}} \\
$x$, $\mathbf{x}$, $\mathbf{X}$ & scalar, vector, and matrix, respectively. \\
$\mathbb{A}$, $\mathbb{C}$ & symbol constellation alphabet, set of complex numbers. \\
$(\cdot)^T$, $(\cdot)^{\dagger}$ & transpose and Hermitian operator. \\
$\text{vec}(\cdot)$ & column-wise vectorization operator. \\
$\text{vec}^{-1}(\cdot)$ & inverse vectorization operator (reshapes to $M \times N$). \\
$\text{diag}\{\mathbf{v}\}$ & diagonal matrix formed from vector $\mathbf{v}$. \\
$\otimes$ & Kronecker product. \\
$\mathcal{CN}(\cdot, \cdot)$ & complex Gaussian distribution. \\
$E[\cdot]$, $\Gamma(\cdot)$ & expectation operator, Gamma function. \\
$\|\cdot\|$, $\arg\min$ & Euclidean norm, argument minimum operator. \\
\addlinespace
\multicolumn{2}{l}{\textit{System and Channel Parameters}} \\
$N_T$, $N_R$ & number of transmit and receive antennas. \\
$M$, $N$ & number of delay bins and Doppler bins. \\
$Q$ & modulation order (number of constellation points). \\
$P$ & number of propagation paths. \\
$h_p$, $A_p$ & complex gain and amplitude of the $p$-th path. \\
$m$, $\Omega$ & Nakagami-$m$ fading parameter, average path power. \\
$l_p$ & integer delay index of the $p$-th path. \\
$k_p$, $\kappa_p$ & integer and fractional Doppler shift of the $p$-th path. \\
\addlinespace
\multicolumn{2}{l}{\textit{Key Matrices}} \\
$\mathbf{I}_K$ & $K \times K$ identity matrix. \\
$\mathbf{F}_M$, $\mathbf{F}_N$ & $M$-point and $N$-point normalized DFT matrices. \\
$\mathbf{G}_{\text{tx}}$, $\mathbf{G}_{\text{rx}}$ & transmit and receive pulse-shaping matrices. \\
$\mathbf{\Pi}$, $\mathbf{\Delta}$ & delay-inducing and Doppler-inducing matrices. \\
$\mathbf{X}_t$ & transmitted DD domain matrix from antenna $t$. \\
$\mathbf{S}_t$ & transmitted TF domain matrix from antenna $t$. \\
$\mathbf{H}_{r,t}$ & time-domain channel matrix for link $(r,t)$. \\
$\mathbf{H}_{\text{eff}}^{(r,t)}$ & effective DD domain channel matrix for link $(r,t)$. \\
$\mathbf{R}_r$ & received TF domain matrix at antenna $r$. \\
$\mathbf{Y}_r$ & received DD domain matrix at antenna $r$. \\
$\mathbf{G}_t$ & effective channel gain matrix for antenna $t$ after MRC. \\
\addlinespace
\multicolumn{2}{l}{\textit{Key Vectors}} \\
$\mathbf{x}_t$ & vectorized DD symbols from antenna $t$. \\
$\hat{\mathbf{x}}_t$ & estimated symbol vector for antenna $t$. \\
$\hat{\mathbf{x}}_{\text{all}}$ & aggregated vector of all estimated symbols. \\
$\mathbf{s}_t$ & transmitted time-domain signal from antenna $t$. \\
$\mathbf{w}_r$ & AWGN vector at receive antenna $r$. \\
$\tilde{\mathbf{w}}_r$ & processed (DD domain) noise vector at antenna $r$. \\
$\mathbf{r}_r$ & received time-domain signal at antenna $r$. \\
$\mathbf{y}_r$ & vectorized DD domain signal at antenna $r$. \\
$\mathbf{n}_t$ & post-combining noise vector for antenna $t$. \\
$\mathbf{z}_t$ & MRC output vector for antenna $t$. \\
\bottomrule
\end{tabularx}
\end{table}

\section{MIMO-OTFS System Model}
\label{sec:mimo_otfs}
\begin{figure*}
    \centering
\includegraphics[width=.8\linewidth]{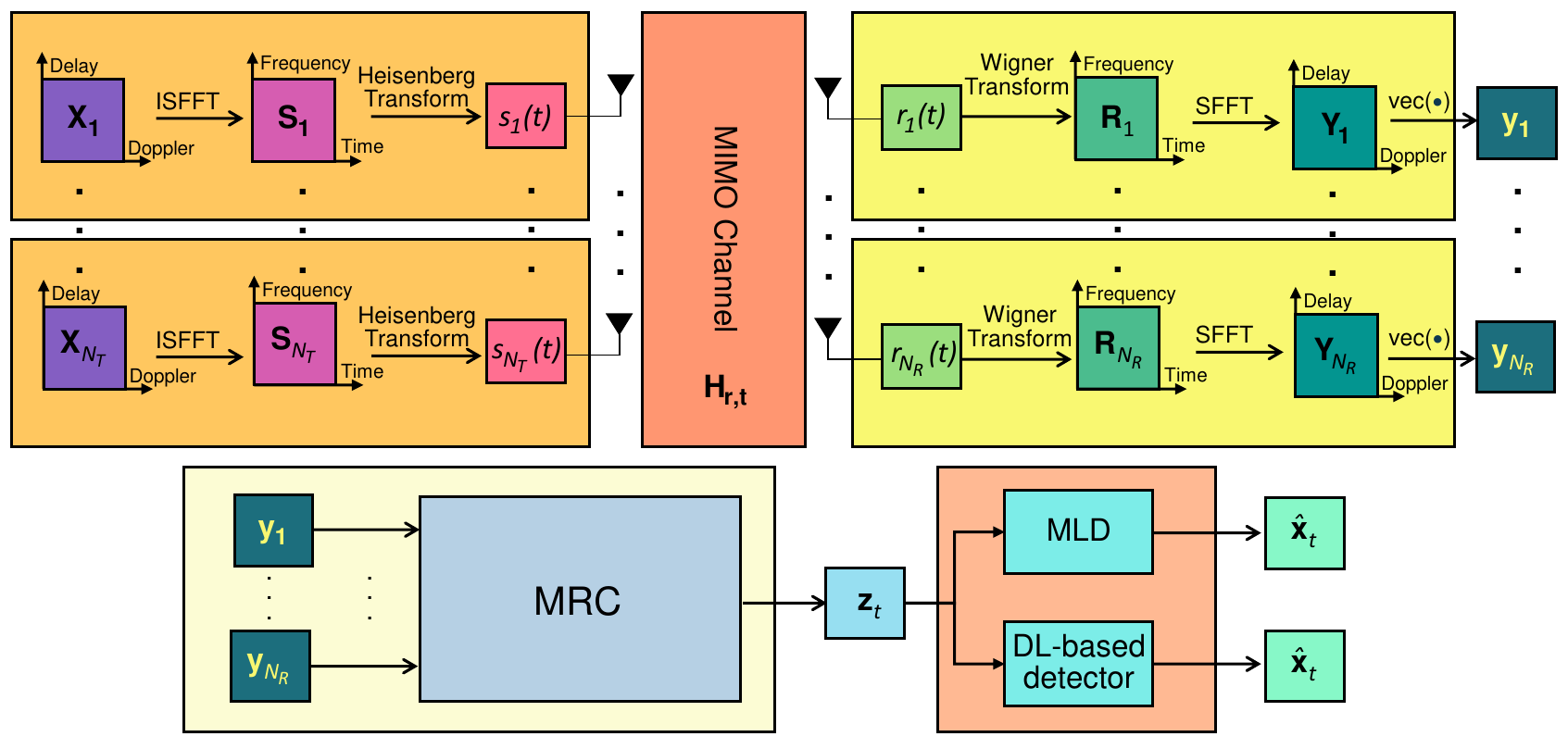}
    \caption{MIMO-OTFS system model.}
    \label{fig:system}
\end{figure*}
This section presents a MIMO-OTFS system with $N_T$ transmit and $N_R$ receive antennas employing spatial multiplexing. Each antenna transmits independent data symbols in the DD domain, and the receiver performs MRC followed by MLD to detect the input symbols. This MIMO-OTFS system model is illustrated in Fig.~\ref{fig:system}.

\subsection{Transmitter}
The symbol generation process consists of two main stages. First, the DD domain matrix $\mathbf{X}_t$ is transformed into the TF domain matrix $\mathbf{S}_t \in \mathbb{C}^{M \times N}$ using the inverse symplectic finite Fourier transform (ISFFT), given by $\mathbf{S}_t = \mathbf{F}_M \mathbf{X}_t \mathbf{F}_N^{\dagger}$. Second, the resulting domain symbol $\mathbf{S}_t$ is converted into the discrete-time signal vector $\mathbf{s}_t \in \mathbb{C}^{MN \times 1}$ using the Heisenberg transform. This transform applies the transmit pulse-shaping matrix $\mathbf{G}_{\text{tx}} \in \mathbb{C}^{M \times M}$ and vectorizes the output, such that $\mathbf{s}_t = \text{vec}(\mathbf{G}_{\text{tx}} \mathbf{S}_t)$. This DD domain matrix $\mathbf{X}_t$ is vectorized as $\mathbf{x}_t = \text{vec}(\mathbf{X}_t)$. For spatial multiplexing, different data matrices $\mathbf{X}_t$ are transmitted from each antenna, i.e., $\mathbf{X}_t \neq \mathbf{X}_{t'}$ for $t \neq t'$.

\subsection{Channel}\label{subsec:channel}
The DD domain channel impulse response for each transmit-receive antenna link $(r,t)$ is characterized by $P_{r,t}$ paths. For convenience of notational simplicity, we omit the $(r,t)$ indices in the subsequent expression as follows \cite{Orthogonal_Time_Frequency_Space_Modulation}:
\begin{equation}
h(\tau, \nu) = \sum_{p=1}^{P} h_p \delta(\tau - \tau_p)\delta(\nu - \nu_p),
\label{eq:channel_mimo}
\end{equation}
where $P$ is the number of propagation paths, and $h_p$, $\tau_p$, and $\nu_p$ are the complex gain, delay, and Doppler shift of the $p$-th path, respectively.
Each path gain $h_p = A_p e^{j\phi_p}$ consists of an amplitude $A_p$ and a phase $\phi_p$. The amplitude $A_p$ is assumed to follow the Nakagami-$m$ distribution, where $m \geq 0.5$ is the fading parameter. The probability density function (PDF) of the amplitude is given by \cite{Nakagami1960}
\begin{equation}
f_A(a) = \frac{2m^m}{\Gamma(m)\Omega^m} a^{2m-1} \exp\left(-\frac{m}{\Omega}a^2\right), \quad a \geq 0,
\label{eq:nakagami}
\end{equation}
where $\Gamma(\cdot)$ is the incomplete Gamma function \cite{Gradstejn_2000} and $\Omega = E[A_p^2]$ represents the average power of the path gain.

In the discrete-time domain, the continuous path parameters (e.g., $\tau_p, \nu_p$) correspond to discrete indices. The time-domain channel matrix $\mathbf{H}_{r,t} \in \mathbb{C}^{MN \times MN}$ for the $(r,t)$ link is constructed from its $P$ paths as
\begin{equation}
\mathbf{H}_{r,t} = \sum_{p=1}^{P} h_p \mathbf{\Pi}^{l_p} \mathbf{\Delta}^{k_p + \kappa_p},
\label{eq:H_mimo}
\end{equation}
where $l_p$ is the integer delay, while $k_p$ and $\kappa_p$ are the integer and fractional Doppler shift, respectively.

\subsection{Receiver}\label{subsec:receiver}
The received time-domain signal vector $\mathbf{r}_r \in \mathbb{C}^{MN \times 1}$ at antenna $r$ is given by the superposition of signals from all $N_T$ transmit antennas as

\begin{equation}
\mathbf{r}_r = \sum_{t=1}^{N_T} \mathbf{H}_{r,t} \mathbf{s}_t + \mathbf{w}_r,
\label{eq:r_mimo}
\end{equation}
where $\mathbf{w}_r \sim \mathcal{CN}(\mathbf{0}, \sigma^2\mathbf{I}_{MN})$ is the additive white Gaussian noise (AWGN) vector.

First, the Wigner transform is applied using the receiver pulse $\mathbf{G}_{\text{rx}}$ to obtain the TF domain grid $\mathbf{R}_r \in \mathbb{C}^{M \times N}$ via the relation $\mathbf{R}_r = \mathbf{G}_{\text{rx}} \, \text{vec}^{-1}(\mathbf{r}_r)$, where $\text{vec}^{-1}(\cdot)$ is the inverse vectorization operator that reshapes the $MN \times 1$ vector into an $M \times N$ matrix.
Second, the symplectic finite Fourier transform (SFFT) is applied to the TF symbol $\mathbf{R}_r$ to obtain the DD domain matrix $\mathbf{Y}_r \in \mathbb{C}^{M \times N}$ is given by
\begin{equation}
\mathbf{Y}_r = \mathbf{F}_M^{\dagger} \mathbf{R}_r \mathbf{F}_N.
\label{eq:sfft_mimo}
\end{equation}
The final received DD domain vector $\mathbf{y}_r \in \mathbb{C}^{MN \times 1}$ is obtained by vectorizing this matrix as $\mathbf{y}_r = \text{vec}(\mathbf{Y}_r)$.
The DD domain input-output relationship is defined as follows:
\begin{equation}
\mathbf{y}_r = \sum_{t=1}^{N_T} \mathbf{H}_{\text{eff}}^{(r,t)} \mathbf{x}_t + \tilde{\mathbf{w}}_r,
\label{eq:y_mimo}
\end{equation}
where the combined vectorized receiver operation is $\left(\mathbf{F}_N \otimes (\mathbf{F}_M^{\dagger} \mathbf{G}_{\text{rx}}) \right)$. The processed noise is $\tilde{\mathbf{w}}_r = \left(\mathbf{F}_N \otimes (\mathbf{F}_M^{\dagger} \mathbf{G}_{\text{rx}}) \right) \mathbf{w}_r$, and the effective DD channel matrix $\mathbf{H}_{\text{eff}}^{(r,t)} \in \mathbb{C}^{MN \times MN}$ is defined as
\begin{equation}
\mathbf{H}_{\text{eff}}^{(r,t)} = \left(\mathbf{F}_N \otimes (\mathbf{F}_M^{\dagger} \mathbf{G}_{\text{rx}}) \right) \mathbf{H}_{r,t} \left(\mathbf{F}_N^\dagger \otimes (\mathbf{G}_{\text{tx}} \mathbf{F}_M) \right).
\label{eq:H_eff_mimo}
\end{equation}

\subsection{MRC-based MLD}\label{subsec:detection}
The proposed receiver is designed as a cascaded two-stage process. In the first stage, MRC is employed to coherently combine the signals from all $N_R$ receive antennas, thereby enhancing the desired signal power and improving the overall signal-to-noise ratio (SNR). The final SD is completed by inserting the symbol vectors from the MRC output into an MLD.

\subsubsection{MRC}
First, the receiver applies MRC for each target transmit antenna $t$. The MRC output vector $\mathbf{z}_t \in \mathbb{C}^{MN \times 1}$, which combines the signals from all $N_R$ receive antennas, is given by:
\begin{equation}
\mathbf{z}_t = \sum_{r=1}^{N_R} \left(\mathbf{H}_{\text{eff}}^{(r,t)}\right)^{\dagger} \mathbf{y}_r.
\label{eq:mrc_mimo}
\end{equation}
To reduce computational complexity, the detector is designed under the assumption that the inter-antenna interference (IAI) is negligible after the MRC stage. This assumes that the combining matrices $\left(\mathbf{H}_{\text{eff}}^{(r,t)}\right)^\dagger$ for the desired antenna $t$ are approximately orthogonal to the effective channels from other antennas. Under this simplifying assumption, substituting \eqref{eq:y_mimo} into \eqref{eq:mrc_mimo} yields a decoupled input-output relation for each antenna $t$:
\begin{equation}
\begin{split}
\mathbf{z}_t &= \sum_{r=1}^{N_R} \left(\mathbf{H}_{\text{eff}}^{(r,t)}\right)^{\dagger} \left[ \mathbf{H}_{\text{eff}}^{(r,t)} \mathbf{x}_t + \sum_{\substack{t'=1\\t'\neq t}}^{N_T} \mathbf{H}_{\text{eff}}^{(r,t')} \mathbf{x}_{t'} + \tilde{\mathbf{w}}_r\right] \\
&\approx \underbrace{\left(\sum_{r=1}^{N_R} \left(\mathbf{H}_{\text{eff}}^{(r,t)}\right)^{\dagger} \mathbf{H}_{\text{eff}}^{(r,t)}\right)}_{\mathbf{G}_t} \mathbf{x}_t + \mathbf{n}_t,
\end{split}
\label{eq:mrc_simplified_mimo}
\end{equation}
where $\mathbf{G}_t \in \mathbb{C}^{MN \times MN}$ is the effective channel gain matrix for antenna $t$ after MRC, and $\mathbf{n}_t = \sum_{r=1}^{N_R} \left(\mathbf{H}_{\text{eff}}^{(r,t)}\right)^{\dagger} \tilde{\mathbf{w}}_r$ is the post-combining noise vector.

\subsubsection{MLD}
Based on the simplified model in \eqref{eq:mrc_simplified_mimo}, the complex joint detection problem is decoupled into $N_T$ independent and smaller MLD problems. 

The MLD enables estimation of the symbol vector $\mathbf{x}_t$ for each antenna $t$ by solving:
\begin{equation}
\hat{\mathbf{x}}_t = \arg\min_{\mathbf{x}_t \in \mathbb{A}^{MN}} \left\|\mathbf{z}_t - \mathbf{G}_t \mathbf{x}_t\right\|^2,
\label{eq:ml_decoupled_mimo}
\end{equation}
where $\mathbb{A}$ denotes the constellation alphabet. The final estimate of all transmitted symbols is obtained by collecting these independent estimates: $\hat{\mathbf{x}}_{\text{all}} = [\hat{\mathbf{x}}_1^T, \ldots, \hat{\mathbf{x}}_{N_T}^T]^T$.

\section{Neural network-based signal detection}
\label{sec:nn_signal_det}

In this study, three different NN architectures (MLP, CNN, and ResNet) were employed for SD, and their BER performances were compared with each other as well as with the MLD-based detector. For each architecture, models were trained at an SNR level of 8~dB and subsequently tested at 0, 4, 8, 12, and 16~dB. The data used for training were not included in the test phase to ensure fair evaluation and prevent overfitting. The comprehensive procedure for generating this synthetic dataset, encompassing the OTFS modulation and channel simulation steps, is systematically outlined in Algorithm~\ref{alg:data-gen}. Subsequently, the raw samples undergo standardization and cross-validation splitting procedures as defined in Algorithm~\ref{alg:preproc} to prepare the inputs for the NNs. This setup enables the assessment of model robustness across varying noise levels. The experiments were conducted under Nakagami fading conditions with \( m = 1 \) and \( m = 2 \). For both fading parameters, tests were performed in SISO and MIMO (\( N_T = 2, N_R = 2 \)) configurations to comprehensively evaluate detection performance.
\begin{algorithm}
\caption{OTFS-based training data generation.}
\label{alg:data-gen}
\DontPrintSemicolon
\SetAlgoSkip{smallskip}   
\SetInd{0.5em}{0.5em}    
\setlength{\algomargin}{0.5em} 
\small 
\KwIn{$M,N, N_T,N_R, M_Q, \text{SNR}_{\text{train}}$, number of frames $N_{\text{frm}}$}
\KwOut{Raw dataset $\mathcal{D}$ for preprocessing (Alg.~\ref{alg:preproc})}
$\mathcal{D} \leftarrow \emptyset$\;
\For{$i = 1$ \KwTo $N_{\text{frm}}$}{
    \tcc{1) Bit generation and DD-domain mapping}
    Generate random bits: $b_i$\;
    QAM mapping: $X_t^{(i)} \in \mathbb{C}^{M \times N}$, for $t=1,\dots,N_T$\;
    \tcc{2) OTFS modulation}
    \For{$t = 1$ \KwTo $N_T$}{
        $S_t^{(i)} = F_M X_t^{(i)} F_N^{\dagger}$ \tcp*{ISFFT}
        $s_t^{(i)} = \operatorname{vec}(G_{\text{tx}} S_t^{(i)})$ \tcp*{Heisenberg}
    }
    \tcc{3) Nakagami-$m$ channel and noise}
    \For{$(r,t)$, $r=1,\dots,N_R$, $t=1,\dots,N_T$}{
        Draw $A_p$ from Nakagami-$m$; draw phases $\phi_p$\;
        $h_p = A_p e^{j\phi_p}$\;
        $H_{r,t}^{(i)} = \sum_{p=1}^{P} h_p \Pi_{l_p} \Delta_{k_p + \kappa_p}$\;
    }
    \For{$r = 1$ \KwTo $N_R$}{
        $r_r^{(i)} = \sum_{t=1}^{N_T} H_{r,t}^{(i)} s_t^{(i)} + w_r^{(i)}$, \quad $w_r^{(i)} \sim \mathcal{CN}(0,\sigma^2 I)$\;
    }
    \tcc{4) OTFS demodulation}
    \For{$r = 1$ \KwTo $N_R$}{
        $R_r^{(i)} = G_{\text{rx}} \operatorname{vec}^{-1}(r_r^{(i)})$ \tcp*{Wigner}
        $Y_r^{(i)} = F_M^{\dagger} R_r^{(i)} F_N$ \tcp*{SFFT}
        $y_r^{(i)} = \operatorname{vec}(Y_r^{(i)})$\;
    }
    \tcc{5) MRC combining}
    \For{$(r,t)$}{
        $H_{\text{eff}}^{(r,t,i)} = (F_N \otimes (F_M^{\dagger} G_{\text{rx}}))\, H_{r,t}^{(i)}\,(F_N^{\dagger} \otimes (G_{\text{tx}} F_M))$\;
    }
    \For{$t = 1$ \KwTo $N_T$}{
        $z_t^{(i)} = \sum_{r=1}^{N_R} \big(H_{\text{eff}}^{(r,t,i)}\big)^{\dagger} y_r^{(i)} \in \mathbb{C}^{MN}$\;
    }
    \tcc{6) DL input/label generation}
    \For{$t = 1$ \KwTo $N_T$}{
        \For{$k = 1$ \KwTo $MN$}{
            $u_{i,t,k} = \begin{bmatrix} \Re\{ z_{t,k}^{(i)} \} \\ \Im\{ z_{t,k}^{(i)} \} \end{bmatrix} \in \mathbb{R}^{2}$\;
            Convert $x_{t,k}^{(i)}$ to bits and class index: $y_{i,t,k} \in \{0,1,\dots,Q-1\}$\;
            Add sample: $\mathcal{D} \leftarrow \mathcal{D} \cup \{(u_{i,t,k}, y_{i,t,k})\}$\;
        }
    }
}
\end{algorithm}
Although the ultimate performance metric of interest in this study is the bit error rate (BER), the proposed deep learning models are trained at the symbol level. Specifically, each QAM symbol is formulated as a multi-class classification target, where the neural networks predict one of \( Q \) constellation classes (e.g., \( Q = 4 \) for 4-QAM). Accordingly, the sparse categorical cross-entropy loss is employed during training. During the testing phase, the predicted symbol classes are deterministically mapped back to their corresponding bit representations, and BER is computed by comparing the recovered bits with the transmitted ones. This approach ensures efficient and stable training while preserving an accurate bit-level performance evaluation consistent with conventional communication metrics.

\begin{algorithm}
\caption{Preprocessing and splitting.}
\label{alg:preproc}
\DontPrintSemicolon
\SetAlgoSkip{smallskip}
\SetInd{0.5em}{0.5em}
\setlength{\algomargin}{0.5em}
\small
\KwIn{Raw dataset $\mathcal{D}$ from Alg.~\ref{alg:data-gen}}
\KwOut{Standardized samples and 5-fold splits $\{\mathcal{D}^{(f)}_{\text{train}}, \mathcal{D}^{(f)}_{\text{val}}\}_{f=1}^{5}$, scaling parameters $(\mu,\sigma)$}
Compute mean and variance from training data only:\;
$\mu = \mathbb{E}[u], \quad \sigma^2 = \operatorname{Var}[u]$\;
Standardize all samples:\;
$\tilde{u} = (u - \mu) / \sigma$\;
Perform 5-fold cross-validation to obtain training/validation splits used in Alg.~\ref{alg:train}:\;
$\{\mathcal{D}^{(f)}_{\text{train}}, \mathcal{D}^{(f)}_{\text{val}}\}_{f=1}^{5}$\;
\end{algorithm}
\subsection{Multi-Layer Perceptron}\label{sec:sub_mlp}

MLP is one of the most fundamental forms of artificial NNs and typically consists of fully connected layers\cite{popescu2009multilayer}. MLP is widely used, especially in classical machine learning problems and as a core component of DL models. In an MLP with three main layers, the input layer contains the raw data received by the model, while the hidden layers transform the input data through weighted connections and activation functions to extract features. Finally, the output layer produces the model’s final predictions and can use different activation functions depending on the type of problem.

\begin{figure*}
    \centering \includegraphics[width=0.7\textwidth]{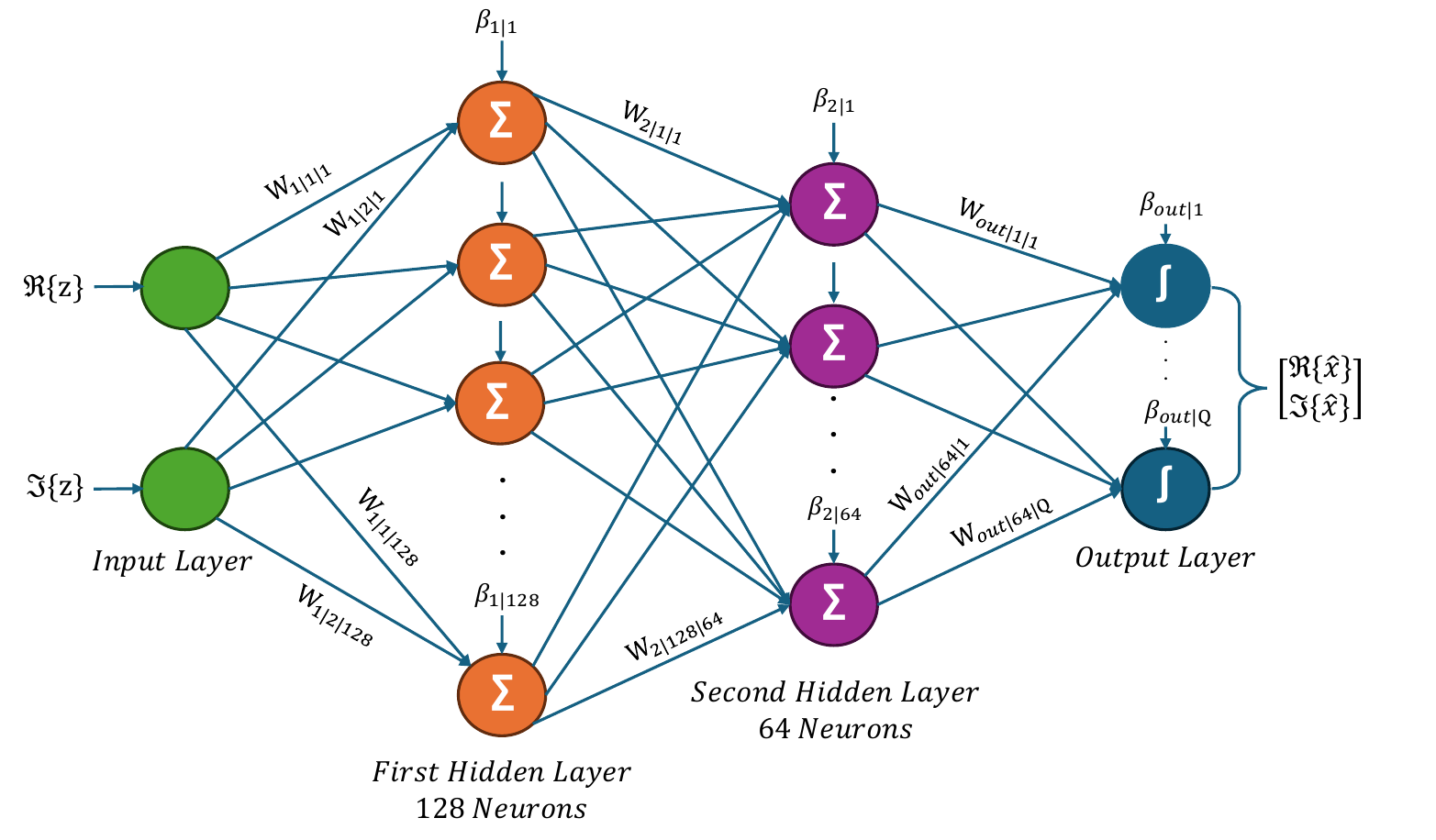}
    \caption{MLP architecture.}
    \label{fig:mlp}
\end{figure*}

The learning process of MLP is based on forward propagation and backpropagation algorithms. In the forward propagation stage, the input data \( x \) is multiplied by the weight matrix \( W \), and the expression \(z = Wx + b\) is computed, followed by applying an activation function \(a = f(z).\) This process continues through the layers, forming the model's predictions. 

In the backpropagation stage, the error between the model's output and the actual value, \( L(y, \hat{y}) \), is calculated, and weights are updated using derivatives. This updating process is performed based on the gradient of the loss function, \( \nabla L \). The weight update is carried out as shown in \eqref{eq:weight_update},

\begin{equation}
W_{\text{new}} = W_{\text{old}} - \eta \nabla L,
\label{eq:weight_update}
\end{equation}
where \( \eta \) is the learning rate.

The MLP structure used in this study is shown in Fig. \ref{fig:mlp}. The MLP used in this study consists of two hidden layers. The first hidden layer has 128 neurons, while the second hidden layer has 64 neurons. The ReLU function was used as the activation function in the hidden layers, whereas the softmax function was used in the output layer for signal prediction. The Adam optimizer was employed for model optimization, and the batch size was set to 4096. The model was trained for 50 epochs.

\subsection{Convolutional Neural Networks}\label{sec:sub_cnn}
\begin{figure*}
    \centering
    \includegraphics[width=.75\textwidth]{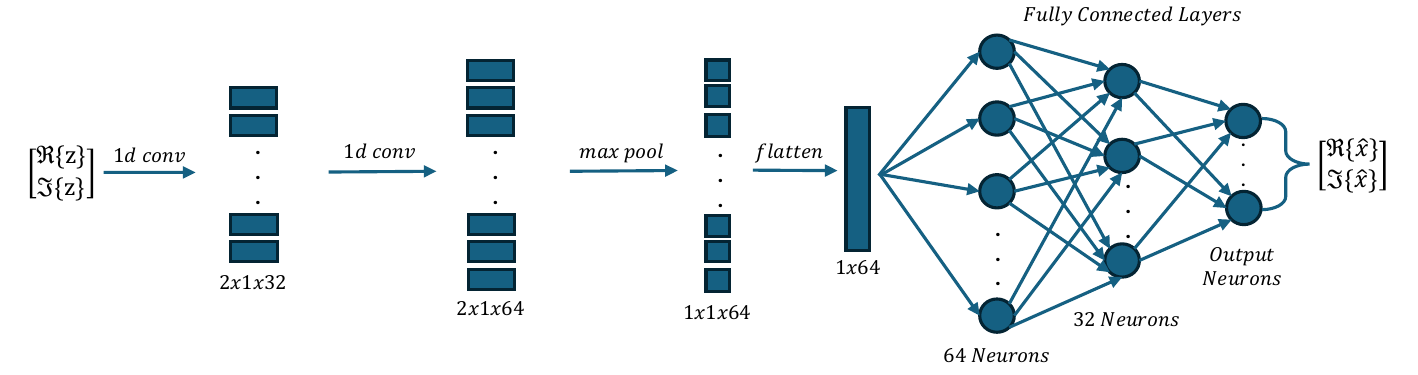}
    \caption{CNN architecture.}
    \label{fig:cnn}
\end{figure*}
CNNs are a type of NN frequently used in image processing and DL problems \cite{o2015introduction}. CNNs utilize convolutional layers to capture local features in the input data. Their fundamental components include convolutional layers, activation functions, pooling layers, and fully connected layers. CNN models are highly effective in extracting meaningful features by filtering the input data at different levels.

A CNN model typically works by first processing the input data with different filters to generate feature maps, allowing convolutional layers to learn weights and extract important features. Then, pooling layers are used to reduce the model's size and lower computational cost. Pooling is usually performed using max or average pooling methods. In the final stage, fully connected layers are used to perform the final classification or regression predictions.

The CNN structure used in this study is shown in Fig. \ref{fig:cnn}. The CNN model consists of two convolutional layers. The first convolutional layer has 32 filters, while the second convolutional layer has 64 filters. Both layers use the ReLU activation function. A max pooling operation is applied after the second convolutional layer. The extracted features are then passed through two fully connected layers with 128 and 64 neurons, respectively. The final output layer has 4 neurons and utilizes the softmax activation function for classification. For a fair comparison, the Adam optimizer was used for the CNN, just like in the MLP, and the batch size was set to 4096. It was trained for 50 epochs.

\subsection{Residual Networks}\label{sec:sub_resnet}

ResNet is a family of models developed to address the vanishing gradient problem encountered in training deep NNs. Proposed by He et al. in 2016, ResNet enhances the trainability of deep networks by utilizing residual connections (skip connections) \cite{he2016deep}. These connections allow the input to bypass certain layers, enabling better learning. Different versions, such as ResNet-50, ResNet-101, and ResNet-152, are commonly used for tasks like image classification, object detection, and segmentation. In particular, Faster R-CNN ResNet50 FPN is a widely used architecture for object detection.

\begin{figure*}[!t]
    \centering
    \includegraphics[width=0.9\textwidth]{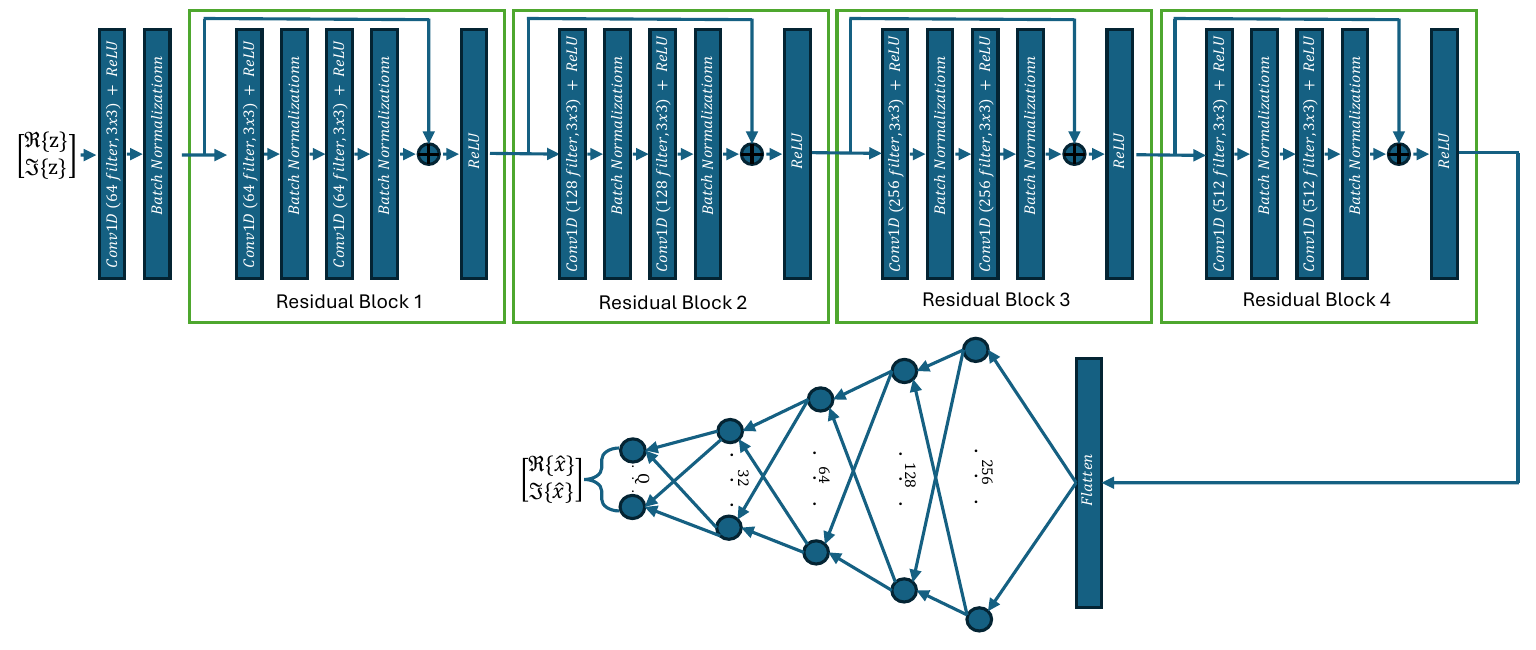}
    \caption{ResNet architecture.}
    \label{fig:resnet}
\end{figure*}

The ResNet architecture used in this study, as shown in Fig. \ref{fig:resnet}, starts with an initial convolutional layer following the input layer. Subsequently, four residual blocks containing 64, 128, 256, and 512 filters are applied. Each residual block consists of two convolutional layers and a shortcut connection. These blocks facilitate the direct transfer of weights, enhancing the model's learning capacity.

The features extracted from the residual blocks are flattened before being passed to the fully connected layers. The fully connected layers of the model contain 256, 128, 64, and 32 neurons, respectively. The final layer has 4 neurons and performs classification using the softmax activation function.

The Adam optimizer was used for model optimization. To maintain fair comparison conditions, the batch size was set to 4096 and the training process was conducted for 50 epochs, consistent with the MLP and CNN models. The large batch size was chosen to reduce update frequency and simplify computation, and even under these settings, the model achieved sufficient performance.

The optimization of network parameters is achieved through the iterative training process detailed in Algorithm \ref{alg:train}, where the sparse categorical cross-entropy loss function was employed during the update steps. This loss is particularly suitable for classification problems where the target labels are provided as integer indices rather than one-hot encoded vectors. In this study, the networks were trained to predict $Q$ distinct symbol classes, where $Q$ represents the modulation order; specifically, $Q=4$ for 4-QAM and $Q=16$ for 16-QAM configurations. This approach allows the detector to generalize across different constellation sizes while maintaining high classification accuracy. 

The loss for a single training sample is defined as

\begin{equation}
L(y, \hat{y}) = - \log \left( \frac{\exp(\hat{y}{y})}{\sum\limits{q=1}^{Q} \exp(\hat{y}_{q})} \right),
\label{eq:scce}
\end{equation}
where \( \hat{y}_q \) denotes the predicted logit for class \( q \) and \( y \) is the true class index.  
This function computes the negative log-likelihood of the true class under the probability distribution obtained via the softmax operation.  

The average loss over a mini-batch is given by

\begin{equation}
\mathcal{L} = \frac{1}{B} \sum_{i=1}^{B} L(y_i, \hat{y}_i),
\end{equation}
where \( B \) is the batch size.  
The use of sparse categorical cross-entropy reduces memory consumption since it directly accepts class indices instead of one-hot encoded vectors, 
resulting in improved computational efficiency, particularly for large-scale datasets.

\section{Complexity Analysis}
\label{sec:complexity}
In this section, we provide a detailed analysis of the computational complexity associated with the SD schemes in the MIMO-OTFS system. We evaluate and compare the computational overhead of the conventional MRC-based MLD against the proposed DL-based architectures (MLP, CNN, and ResNet). This comparison, presented in terms of big-O $(\mathcal{O})$ notation and the number of real multiplications (RMs), highlights the trade-off between detection performance and computational efficiency, which is critical for practical implementation in future wireless networks.

\subsection{Complexity of SD}\label{sec:complexity_mrc_ml}

The computational complexity of the MRC-based MLD is analyzed by examining all processing stages from received signal demodulation to final SD.

\textit{Receiver Demodulation:} At each receive antenna $r$, the Wigner transform and SFFT operation $\mathbf{Y}_r = \mathbf{F}_M^{\dagger} \mathbf{R}_r \mathbf{F}_N$ require $\mathcal{O}(MN \log(MN))$ operations using FFT algorithms. For $N_R$ receive antennas, this results in $\mathcal{O}(N_R \cdot MN \log(MN))$ complexity. In terms of RMs, this stage requires $4 N_R MN \log(MN)$ operations.

\textit{Effective Channel Matrix Computation:} The computation of $\mathbf{H}_{\text{eff}}^{(r,t)}$ in \eqref{eq:H_eff_mimo} involves Kronecker products and matrix multiplications of size $MN \times MN$ for each transmit-receive antenna pair $(r,t)$. This requires $\mathcal{O}(N_T N_R (MN)^3)$ operations, corresponding to $4 N_T N_R (MN)^3$ RMs.

\textit{MRC:} This stage involves two main computations for each of the $N_T$ transmit antennas. First, the effective channel gain matrix $\mathbf{G}_t$ is computed as defined in \eqref{eq:mrc_simplified_mimo}, which involves $N_R$ matrix-matrix multiplications $(\mathbf{H}_{\text{eff}}^{(r,t)})^{\dagger} \mathbf{H}_{\text{eff}}^{(r,t)}$ of size $MN \times MN$. This results in a complexity of $\mathcal{O}(N_R (MN)^3)$ per transmit antenna, totaling $\mathcal{O}(N_T N_R (MN)^3)$ for all antennas, which corresponds to $4 N_T N_R (MN)^3$ RMs. Second, the MRC output vector $\mathbf{z}_t$ is computed as in \eqref{eq:mrc_mimo}, requiring $N_R$ matrix-vector multiplications with complexity $\mathcal{O}(N_R (MN)^2)$ per antenna, totaling $\mathcal{O}(N_T N_R (MN)^2)$ for all antennas, corresponding to $4 N_T N_R (MN)^2$ RMs. Therefore, the effective channel computation and MRC preprocessing stages have a total complexity of $\mathcal{O}(N_T N_R (MN)^3)$ or $8 N_T N_R (MN)^3 + 4 N_T N_R (MN)^2$ RMs.

\textit{MLD:} This stage performs the final symbol estimation by operating on the MRC output vectors $\mathbf{z}_t$ for each transmit antenna. In practical implementations, this is typically performed as symbol-by-symbol detection (i.e., assuming $\mathbf{G}_t$ is (approximately) diagonal so that decisions decouple across symbol positions). For each symbol position $k$, the metric computation $|z_{t,k} - g_{t,k} x|^2$ requires 6 RMs (4 RMs for the complex multiplication $g_{t,k}x$ and 2 RMs for $|a+jb|^2=a^2+b^2$). For $N_T$ transmit antennas with $MN$ symbols each, the total MLD complexity is $6 N_T Q MN$ RMs.

\textit{Overall Complexity:} The total complexity of the MRC-MLD is given by
$C_{\text{MRC-ML}} = 4 N_R MN \log(MN) + 8 N_T N_R (MN)^3 + 4 N_T N_R (MN)^2 + 6 N_T Q MN$,
where the cubic term $8 N_T N_R (MN)^3$ from the effective channel computation and MRC preprocessing stages dominates the overall computational cost.

\subsection{Complexity of DL-based SD}

In this study, various deep networks were used, each having different complexities based on the hidden layers, convolutional layers, and skip connections in ResNet architectures. It is crucial to emphasize that the computational complexity of these DL-based methods is predominantly determined by a fixed forward propagation through a pretrained network, rather than the iterative metric search required by traditional MLD. While the following analytical derivations are specifically presented for the 4-QAM scenario ($Q=4$), the total RM counts remain remarkably stable when extending the framework to higher-order modulations such as 16-QAM ($Q=16$). This stability arises because the overall complexity is primarily dominated by the OTFS block size $MN$ within the input and hidden layers, whereas changing the constellation size $Q$ only necessitates a minor adjustment in the final output layer's dimensions, contributing a negligible fraction to the total RM count. Consequently, the proposed DL-based approaches ensure a predictable and scalable inference cost, avoiding the constellation-dependent complexity growth inherent to conventional detection schemes.

The MLP architecture employed in this study consists exclusively of fully connected layers.
The computational complexity of an MLP is expressed as
\begin{equation}
\sum_{j=1}^J \mathcal{O} \left( d_{j-1} d_j \right),
\label{eq:mlp_comp}
\end{equation}
where $j$ denotes the layer index, $J$ is the total number of layers, and $d_{j-1}$ and $d_j$ denote the number of neurons in the previous and current layers, respectively.

When summing the cost of all layers, the general expression becomes
\begin{equation}
\mathcal{O}\bigl((MN)d_{1}
\;+\; d_{1} d_{2}
\;+\; \dots
\;+\; d_{J-1} d_{J}\bigr).
\label{eq:mlp_long_comp}
\end{equation}
Since the hidden layer dimensions are fixed and relatively small, the dominant contribution arises from the first layer, and thus the asymptotic complexity of the MLP scales linearly with the block size as $\mathcal{O}(MN)$.

Unlike the MRC-based detector, which operates on complex-valued signals, the MLP processes real-valued feature representations obtained from the MRC output.
Accordingly, all computations in the MLP are performed using real-valued arithmetic, and each weight multiplication in a fully connected layer corresponds to a single real multiplication.

Therefore, the RM complexity of the MLP can be written in its classical form as
\begin{equation}
C_{\text{MLP}}
=
\sum_{j=1}^J d_{j-1} d_j.
\end{equation}
For the specific MLP configuration used in this study, consisting of an input layer of size $MN$, two hidden layers with 128 and 64 neurons, and a 4 neurons output layer, this expression simplifies to
$
C_{\text{MLP}} = 128\times MN + 128 \times 64 + 64 \times 4 = 128\times MN + 8448
$.

The computational complexity of the CNN model consists of the convolutional layers followed by
a fully connected classification stage.
The overall complexity can be expressed as
\begin{equation}
\sum_{i=1}^I \mathcal{O} \left( L_i K_i F_{i-1} F_i \right)
+
\sum_{j=1}^J \mathcal{O} \left( d_{j-1} d_j \right),
\label{eq:cnn_comp}
\end{equation}
where $MN$ denotes the OTFS block size, $L_i$ represents the output length of the $i$-th
convolutional layer (which equals $MN$ under stride-1 and same-padding convolutions),
$K_i$ is the kernel size, and $F_{i-1}$ and $F_i$ denote the number of input and output channels, respectively.

Similar to the MLP case, the CNN operates entirely on real-valued feature representations.
Each convolution output sample requires $K_i F_{i-1}$ real-valued multiplications per output channel.
Accordingly, the RM complexity of the CNN can be written in classical form as
\begin{equation}
C_{\text{CNN}}
=
\sum_{i=1}^I L_i K_i F_{i-1} F_i
+
\sum_{j=1}^J d_{j-1} d_j.
\end{equation}

For the CNN architecture employed in this study, the input consists of two real-valued channels,
corresponding to $\begin{bmatrix}
\Re\{ z_{t,k}\}, \Im\{ z_{t,k}\}
\end{bmatrix}^T$.
The network comprises two 1D convolutional layers with kernel size $K=3$ and channel dimensions
$F_0=2$, $F_1=32$, and $F_2=64$.
Assuming stride-1 and same-padding convolutions, the convolutional RM cost becomes $MN \times 3 \times 2 \times 32
+
MN \times 3 \times 32 \times 64
=
6336MN$.

Following the convolutional layers, a max-pooling operation with pooling factor $2$ is applied,
which reduces the feature length to $MN/2$ without introducing additional multiplications.
The resulting feature map is flattened and passed to a fully connected (FC) classifier with layer sizes
$32MN \!\to\! 64 \!\to\! 32 \!\to\! 4$.
The corresponding RM complexity of the FC stage is therefore $2048MN + 2176$.

Combining all components, the total RM complexity of the CNN detector is given by $C_{\text{CNN}} = 8384MN + 2176$,
which scales linearly with the OTFS block size.
Since all kernel sizes, channel dimensions, and layer widths are fixed,
the asymptotic complexity of the CNN model is $\mathcal{O}(MN)$.
The ResNet architecture extends the CNN model by incorporating residual blocks with skip
connections, enabling deeper representations while preserving gradient flow.
Based on the architecture shown in Fig.~\ref{fig:resnet}, the input consists of two real-valued
channels $\begin{bmatrix}
\Re\{ z_{t,k} \}, \Im\{ z_{t,k} \}
\end{bmatrix}^T$ of length $MN$, which are first processed by
an initial 1D convolutional layer with kernel size $K_0=3$ and $F_1=64$ output channels.
This initial convolution results in an RM cost of $384MN$.

Following the initial convolution, the network comprises $L=4$ residual blocks with channel
dimensions $64$, $128$, $256$, and $512$.
Each residual block contains two 1D convolutional layers with kernel size $K=3$.
Stride-$2$ convolution is applied only at the beginning of each residual block to reduce the
temporal resolution, leading to a progressive reduction of the feature length as $MN/2^l$
at the $l$-th block.
Skip connections involve only element-wise additions and therefore do not contribute to the RM count.

Accordingly, the RM complexity of the convolutional part of the ResNet can be expressed as
\begin{equation}
\begin{split}
C_{\text{conv}}
=
& MN \cdot K_0 F_0 F_1 \\
& + \sum_{l=1}^{L}
\left[
\frac{MN}{2^{l}}
\left(
K F_{l-1} F_l + K F_l^2
\right)
\right],
\end{split}
\end{equation}
where the two terms inside the summation correspond to the two convolutional layers within
each residual block.

For the adopted architecture, the RM costs of the four residual blocks are
$12288MN$, $18432MN$, $36864MN$, and $73728MN$, respectively, yielding a total convolutional
complexity of $C_{\text{conv}} = 141696MN$.

After the final residual block, the feature map is flattened and passed to a FC classifier with layer sizes $512 \!\to\! 256 \!\to\! 128 \!\to\! 64 \!\to\! 32 \!\to\! 4$.
The corresponding RM complexity of the FC stage is $C_{\text{FC}} = 512\!\times\!256 + 256\!\times\!128 + 128\!\times\!64 + 64\!\times\!32 + 32\!\times\!4
= 174208$.

Combining all components, the total RM complexity of the ResNet detector is $C_{\mathrm{ResNet}} = 141696MN + 174208$,
which scales linearly with the OTFS block size.
Since all kernel sizes, channel dimensions, and network depth are fixed,
the asymptotic complexity of the ResNet model is $\mathcal{O}(MN)$.

\subsection{Comparison Between ML and DL-Based Detection}
\label{subsec:ml_vs_dl}

While both ML and DL-based detectors operate on the same MRC-processed signal representation, their computational characteristics differ fundamentally due to the nature of their detection mechanisms. ML detection relies on explicit metric evaluations over the modulation constellation for each transmitted symbol, whereas DL-based detection replaces this iterative search with a fixed forward propagation through a pretrained neural network. This conceptual difference leads to markedly different online computational behaviors.

As shown in Table~\ref{tab:complexity_6g_comprehensive}, the complexity of ML detection scales linearly with the number of transmit antennas and the constellation size, as each symbol position requires repeated metric evaluations over the alphabet set. The antenna configurations are selected based on 6G deployment scenarios where transmit antenna counts can reach up to 512~\cite{10379539}, and the MLD complexity is computed as $6 N_T Q MN$ real multiplications (RMs) using a symbol-wise metric approach. Consequently, higher-order modulations such as 1024-QAM significantly increase the computational burden of ML detection. In contrast, the computational cost of DL-based detectors remains constant once the network architecture is fixed, regardless of the modulation order or antenna configuration. This property enables DL-based detectors to maintain a stable and predictable inference cost across diverse system settings.

Among the considered DL architectures, the MLP and CNN models exhibit particularly favorable complexity profiles, achieving several orders of magnitude lower real-multiplication counts than MLD in high spectral-efficiency regimes. Although the ResNet architecture incurs a higher computational cost due to its deeper structure and residual blocks, its complexity remains bounded and does not scale with the modulation order. As a result, even the most complex DL model considered in this study avoids the exponential or constellation-dependent growth inherent to ML-based detection.

\begin{figure}
    \centering
    \includegraphics[width=1\columnwidth]{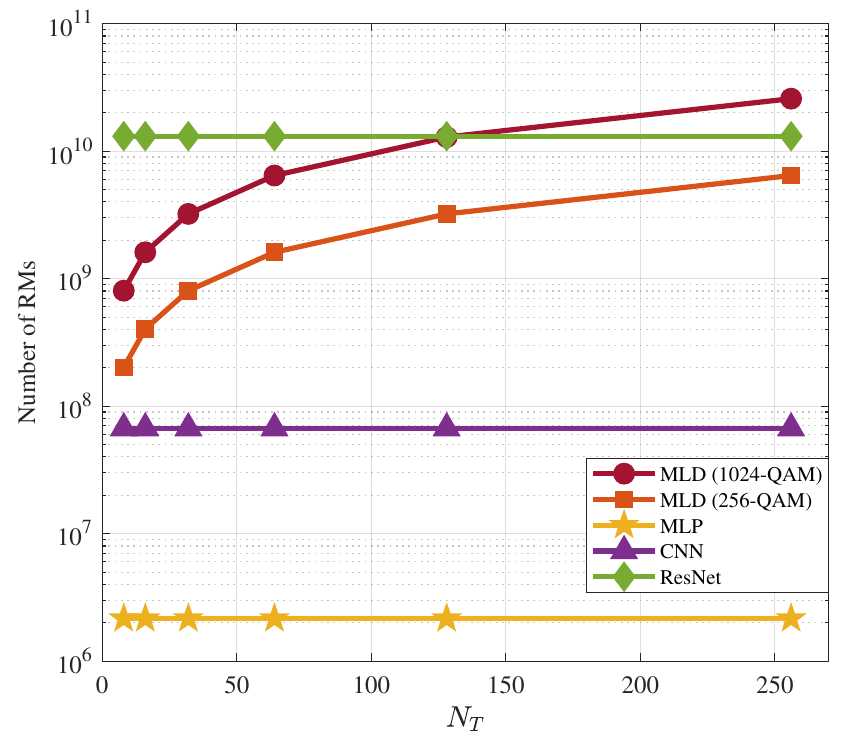}
    \caption{Computational complexity comparison for massive MIMO configurations ($M = 128$, $N = 128$) with 256-QAM and 1024-QAM modulation. MLD complexity scales with $N_T$ while DL-based detectors maintain constant complexity.}
    \label{fig:comparison}
\end{figure}

It is important to note that the reported DL complexities correspond exclusively to the online inference stage. The training phase, which constitutes the most computationally intensive part of deep learning, is performed offline only once and therefore does not affect real-time detection latency. In contrast, MLD incurs its full computational cost repeatedly for every OTFS block. This distinction renders DL-based detectors particularly attractive for real-time and large-scale MIMO systems, where low and consistent detection latency is a critical requirement. The resulting scalability advantage of DL-based detection is further illustrated in Fig.~\ref{fig:comparison}, where ML complexity increases with system dimensions, while DL-based approaches exhibit constant complexity.

\begin{table}[!h]
\caption{Computational complexity for 6G massive MIMO scenarios with $M = 128$, $N = 128$ ($MN = 16{,}384$).}
\label{tab:complexity_6g_comprehensive}
\centering
\footnotesize
\sisetup{scientific-notation=true, round-mode=places, round-precision=2}
\renewcommand{\arraystretch}{.8}
\setlength{\tabcolsep}{3pt}
\begin{tabular}{|c|c|c|c|c|c|}
\hline
\textbf{$N_T$} & \textbf{$Q$} & \textbf{MLD} & \textbf{MLP} & \textbf{CNN} & \textbf{ResNet} \\
\hline
8 & 256 & \(2.01\times 10^8\) & \(2.12\times 10^6\) & \(1.37\times 10^8\) & \(2.32\times 10^{9}\) \\
\hline
8 & 1024 & \(8.05\times 10^8\) & \(2.17\times 10^6\) & \(1.37\times 10^8\) & \(2.32\times 10^{9}\) \\
\hline
16 & 256 & \(4.03\times 10^8\) & \(2.12\times 10^6\) & \(1.37\times 10^8\) & \(2.32\times 10^{9}\) \\
\hline
16 & 1024 & \(1.61\times 10^9\) & \(2.17\times 10^6\) & \(1.37\times 10^8\) & \(2.32\times 10^{9}\) \\
\hline
32 & 256 & \(8.05\times 10^8\) & \(2.12\times 10^6\) & \(1.37\times 10^8\) & \(2.32\times 10^{9}\) \\
\hline
32 & 1024 & \(3.22\times 10^9\) & \(2.17\times 10^6\) & \(1.37\times 10^8\) & \(2.32\times 10^{9}\) \\
\hline
64 & 256 & \(1.61\times 10^9\) & \(2.12\times 10^6\) & \(1.37\times 10^8\) & \(2.32\times 10^{9}\) \\
\hline
64 & 1024 & \(6.44\times 10^9\) & \(2.17\times 10^6\) & \(1.37\times 10^8\) & \(2.32\times 10^{9}\) \\
\hline
128 & 1024 & \(1.29\times 10^{10}\) & \(2.17\times 10^6\) & \(1.37\times 10^8\) & \(2.32\times 10^{9}\) \\
\hline
256 & 1024 & \(2.58\times 10^{10}\) & \(2.17\times 10^6\) & \(1.37\times 10^8\) & \(2.32\times 10^{9}\) \\
\hline
\end{tabular}
\end{table}

\section{Numerical Results and Discussions}
\label{sec:num_res_dis}
This section investigates the performance of the proposed DL-based SD methods for MIMO-OTFS systems. The evaluation includes both SISO and MIMO configurations, and it uses Nakagami-$m$ fading with varying fading parameters ($m=1$ and $m=2$) to assess system robustness at different channel fading severity levels. The proposed MLP, CNN, and ResNet architectures are compared to the conventional MLD detector in terms of BER over SNR values. First, the simulation parameters and system configuration are explained, and then the BER performance results are analyzed in details. The system parameters are summarized in Table~\ref{tab:system_parameters}.

\begin{table}[ht]
\centering
\caption{System parameters.}
\label{tab:system_parameters}
\begin{tabular}{|>{}l|l|}
\hline
\textbf{Parameter} & \textbf{Value} \\ \hline
Carrier frequency ($f_c$ (GHz)) & $4$ \\ \hline
Subcarrier spacing ($\Delta f$ (kHz)) & $15$ \\ \hline
Number of delay bins ($M$) & $64$ \\ \hline
Number of Doppler bins ($N$) & $64$ \\ \hline
Number of transmit antennas ($N_T$) & ${1,2}$ \\ \hline
Number of receive antennas ($N_R$) & ${1,2}$ \\ \hline
Modulation schemes & {4-QAM, 16-QAM}\\ \hline
Channel model & Nakagami-$m$ fading \\ \hline
Fading parameter ($m$) & ${1 \text{ (Rayleigh)},2}$ \\ \hline
Number of delay taps & 9 \\ \hline
Maximum user speed (km/h) & $120$ \\ \hline
Maximum Doppler shift ($f_{d,\text{max}}$ (Hz)) & $\approx 444.4$ \\ \hline
Symbol duration $(T_s$ ($\mu$s)) & $\approx 1.04$ \\ \hline
Detection methods & MLD, MLP, CNN, ResNet \\ \hline
Target symbols per simulation & $10^5$ \\ \hline
\end{tabular}
\end{table}

\subsection{Simulation Parameters}
This section utilizes Monte Carlo techniques to derive simulation results. In the MATLAB environment, the performance of the SISO-OTFS system was evaluated under various channel conditions within a Nakagami-\textit{m} fading channel. Initially, the number of paths, $P$ is set to 9, and the parameters $M$ and $N$ are both chosen as 64. The channel coefficients are randomly generated according to the Nakagami-\( m \) distribution. To evaluate the system's sensitivity to fading effects, the fading parameter \( m \) is set to 1 and 2, respectively. The dataset for the deep networks used in the study was generated using the specified parameters.
\begin{figure*}[!t]
    \centering
    \begin{minipage}{0.5\textwidth}
        \centering
        \includegraphics[width=\linewidth]{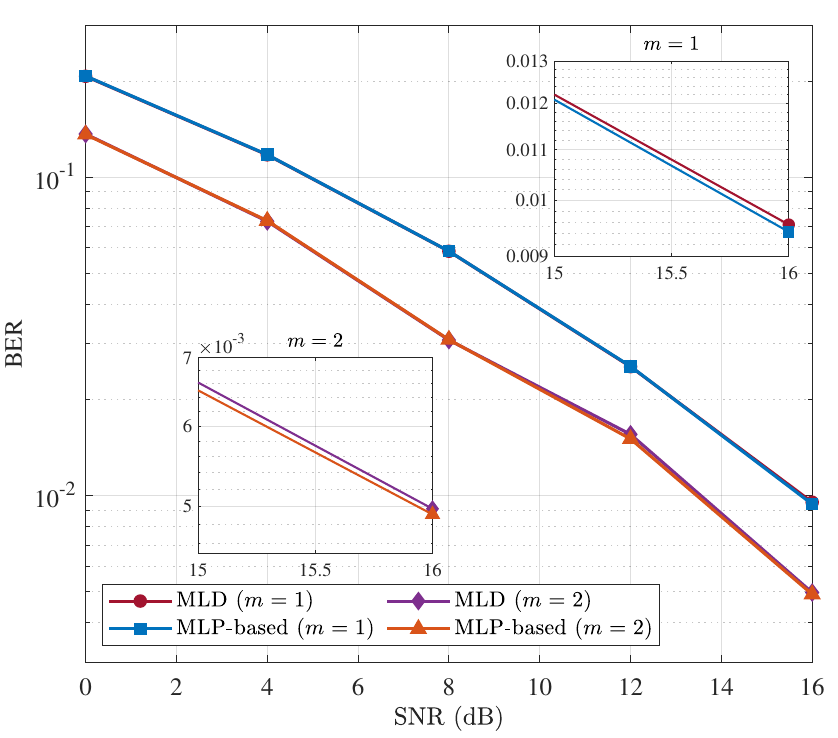}
    \end{minipage}%
    \hfill
    \begin{minipage}{0.5\textwidth}
        \centering
        \includegraphics[width=\linewidth]{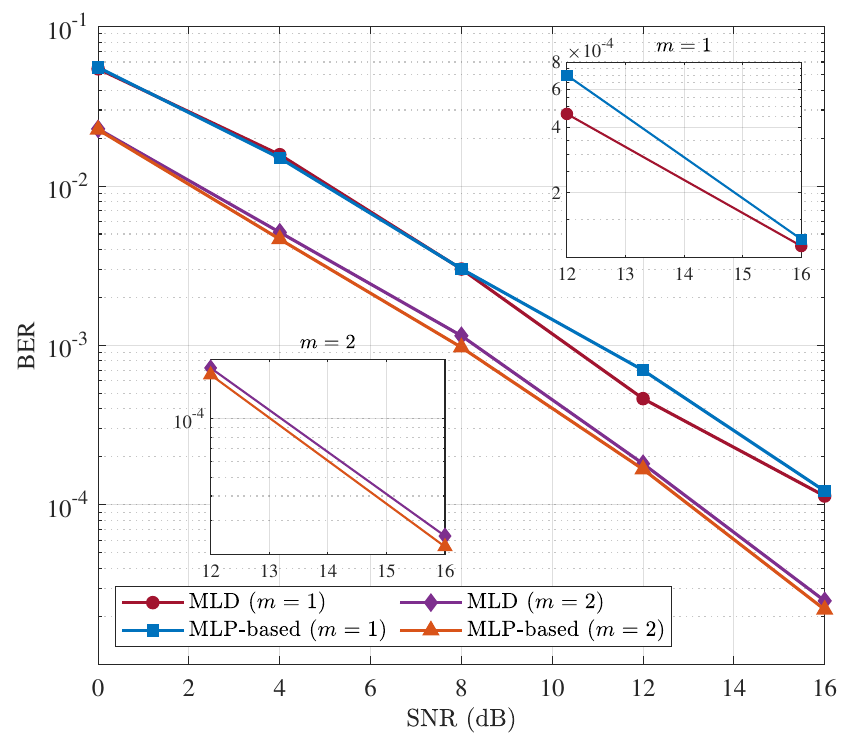}
    \end{minipage}
    \caption{BER results of MLD and MLP-based detector for SISO-OTFS (left) and MIMO-OTFS with $N_T=2$ and $N_R=2$ (right).}
    \label{fig:ber_perf}
\end{figure*}
\begin{figure}
    \centering
    \includegraphics[width=1\linewidth]{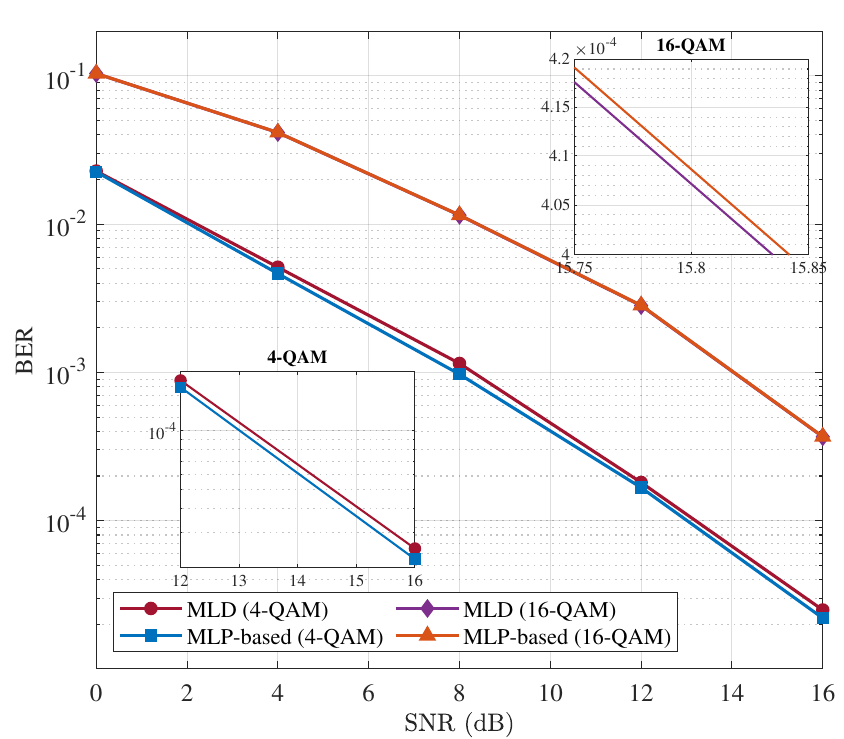}
    \caption{BER results of MLD and MLP-based detector for 4-QAM and 16-QAM.}
    \label{fig:16qam}
\end{figure}

For each network input, the real and imaginary components were provided as two separate input features. 
To ensure statistical reliability, a stratified five-fold cross-validation scheme was employed at an operating SNR of \(8\,\mathrm{dB}\). 
In each fold, 80\% of the data were used for training and 20\% for validation, where the scaling transformation was fitted exclusively on the training subset and subsequently applied to the validation data to avoid data leakage. 
Throughout all NN configurations, identical training conditions were maintained: an initial learning rate of \(10^{-3}\), a maximum of 50 epochs, mini-batches of size 4096, and an early stopping criterion monitoring the validation loss with a patience of ten epochs. 
Additionally, the learning rate was automatically reduced by a factor of 0.5 when no improvement was observed for four consecutive epochs. 

After completing the cross-validation process, the model with the best-performing configuration was retrained using the entire \(8\,\mathrm{dB}\) training dataset. 
It is important to note that these training samples were not included in any of the subsequent test evaluations. The retrained model was then tested under different channel conditions at \(0\), \(4\), \(8\), \(12\), and \(16\,\mathrm{dB}\) SNR levels, following the testing and symbol detection procedure described in Algorithm \ref{alg:test}, to assess its generalization capability. At each SNR level, the BER was calculated based on discrepancies between the predicted and transmitted symbols, and the results were used to evaluate the robustness and generalization performance of all proposed networks.

\begin{algorithm}
\caption{Training of DL model (MLP / CNN / ResNet).}
\label{alg:train}
\DontPrintSemicolon
\SetAlgoSkip{smallskip}
\SetInd{0.5em}{0.5em}
\setlength{\algomargin}{0.5em}
\small
\KwIn{$\{\mathcal{D}^{(f)}_{\text{train}}, \mathcal{D}^{(f)}_{\text{val}}\}_{f=1}^{5}$ from Alg.~\ref{alg:preproc}, max epochs $E_{\max}$, batch size $B$}
\KwOut{Trained parameters $\theta_A^\star$ for each architecture $A$}
\ForEach{architecture $A \in \{\text{MLP}, \text{CNN}, \text{ResNet}\}$}{
    Initialize parameters $\theta_A^{(0)}$\;
    Initialize Adam moment vectors: $m^{(0)} = 0,\ v^{(0)} = 0$\;
    \For{$f = 1$ \KwTo $5$}{
        $e \leftarrow 0$\;
        \While{$e < E_{\max}$ \textbf{and} no early stopping}{
            Split $\mathcal{D}^{(f)}_{\text{train}}$ into minibatches: $\mathcal{B}_1,\dots,\mathcal{B}_J$\;
            \For{$j = 1$ \KwTo $J$}{
                \tcc{1) Forward pass}
                For each $(\tilde{u}_n, y_n) \in \mathcal{B}_j$:
                \Indp
                    $z_n = f_A(\tilde{u}_n; \theta)$ \tcp*{logits}
                    $p_n = \operatorname{softmax}(z_n)$\;
                \Indm
                \tcc{2) Sparse Categorical Cross-Entropy}
                $L_j = \frac{1}{|\mathcal{B}_j|} \sum_{(\tilde{u}_n, y_n) \in \mathcal{B}_j} \left[-\log p_n[y_n]\right]$\;
                \tcc{3) Backprop + Adam}
                Compute gradient: $\nabla_\theta L_j$\;
                $m \leftarrow \beta_1 m + (1-\beta_1)\nabla_\theta L_j$\;
                $v \leftarrow \beta_2 v + (1-\beta_2)(\nabla_\theta L_j)^2$\;
                $\hat{m} = m / (1-\beta_1^{t})$, $\hat{v} = v / (1-\beta_2^{t})$, \quad $t \leftarrow t+1$\;
                Learning rate $\eta_t$ (e.g., ReduceLROnPlateau)\;
                $\theta \leftarrow \theta - \eta_t \, \hat{m} / (\sqrt{\hat{v}} + \epsilon)$\;
            }
            Compute validation loss: $L_{\text{val}}^{(f,e)} = \frac{1}{|\mathcal{D}^{(f)}_{\text{val}}|} \sum_{(\tilde{u},y) \in \mathcal{D}^{(f)}_{\text{val}}} -\log \operatorname{softmax}(f_A(\tilde{u};\theta))[y]$\;
            If $L_{\text{val}}^{(f,e)}$ is smallest so far: $\theta_A^{\text{best}} \leftarrow \theta$\;
            $e \leftarrow e+1$\;
        }
    }
    Set best model: $\theta_A^\star \leftarrow \theta_A^{\text{best}}$\;
}
\end{algorithm}

\begin{algorithm}
\caption{Testing and symbol detection.}
\label{alg:test}
\DontPrintSemicolon
\SetAlgoSkip{smallskip}
\SetInd{0.5em}{0.5em}
\setlength{\algomargin}{0.5em}
\small
\KwIn{Trained model parameters $\theta_A^\star$ from Alg.~\ref{alg:train}, scaling parameters $(\mu,\sigma)$ from Alg.~\ref{alg:preproc}, test SNR set $\{\text{SNR}_{\text{test}}\}$}
\KwOut{Detected QAM symbols $\hat{x}$}
\ForEach{$\text{SNR}_{\text{test}} \in \{0,4,8,12,16\}~\text{dB}$}{
    Repeat OTFS modulation, channel, noise, demodulation and MRC as in Alg.~\ref{alg:data-gen} to obtain $\{z_t^{(\text{test})}\}$\;
    \For{each $z_t^{(\text{test})}$ and each grid index $k$}{
        $\tilde{u}_{t,k}^{(\text{test})} = \begin{bmatrix} \Re\{ z_{t,k}^{(\text{test})} \} \\ \Im\{ z_{t,k}^{(\text{test})} \} \end{bmatrix}$\;
        Standardize using same $\mu,\sigma$\;
        $z_{t,k} = f_A(\tilde{u}_{t,k}^{(\text{test})}; \theta_A^\star)$\;
        $p_{t,k} = \operatorname{softmax}(z_{t,k})$\;
        $\hat{c}_{t,k} = \arg\max_j p_{t,k}[j]$\;
        Convert class $\hat{c}_{t,k}$ to bits and QAM symbols\;
    }
}
\end{algorithm}
\subsection{BER Performance Analysis}

In this section, the BER performance of the MLD is compared with DL-based detection methods. Tables \ref{tab:ber_siso} and \ref{tab:ber_mimo} present the BER performance results obtained at different SNR values for $m=1$ and $m=2$. Additionally, these performance results are visualized in Fig. \ref{fig:ber_perf}.

\begin{table*}
\centering
\caption{BER performance comparison for SISO system ($N_T=1$, $N_R=1$) with $m=1$ and $m=2$.}
\label{tab:ber_siso}
\small
\renewcommand{\arraystretch}{1.0}
\setlength{\tabcolsep}{4pt}
\begin{tabular}{|l|ccccc|ccccc|}
\hline
\multirow{2}{*}{\textbf{Method}} 
& \multicolumn{5}{c|}{$m=1$ (Rayleigh fading)} 
& \multicolumn{5}{c|}{$m=2$} \\ 
\cline{2-11}
& 0 dB & 4 dB & 8 dB & 12 dB & 16 dB
& 0 dB & 4 dB & 8 dB & 12 dB & 16 dB \\
\hline
\textbf{MLD} 
& 0.207282 & 0.117487 & 0.058562 & 0.025438 & 0.009554
& 0.136612 & 0.072703 & 0.030839 & 0.015597 & 0.004976 \\
\textbf{MLP}        
& 0.207712 & 0.117672 & 0.058598 & 0.025505 & 0.009425
& 0.136162 & 0.073009 & 0.030915 & 0.015096 & 0.004913 \\
\textbf{CNN}        
& 0.207655 & 0.117662 & 0.058634 & 0.025508 & 0.009418
& 0.136264 & 0.073044 & 0.030943 & 0.015106 & 0.004924 \\
\textbf{ResNet}     
& 0.207694 & 0.117689 & 0.058640 & 0.025494 & 0.009426
& 0.136216 & 0.073034 & 0.030946 & 0.015093 & 0.004936 \\
\hline
\end{tabular}
\end{table*}

\begin{table*}
\centering
\caption{BER performance comparison for MIMO system ($N_T=2$, $N_R=2$) with $m=1$ and $m=2$.}
\label{tab:ber_mimo}
\small
\renewcommand{\arraystretch}{1.0}
\setlength{\tabcolsep}{4pt}
\begin{tabular}{|l|ccccc|ccccc|}
\hline
\multirow{2}{*}{\textbf{Method}} 
& \multicolumn{5}{c|}{$m=1$ (Rayleigh fading)} 
& \multicolumn{5}{c|}{$m=2$} \\ 
\cline{2-11}
& 0 dB & 4 dB & 8 dB & 12 dB & 16 dB
& 0 dB & 4 dB & 8 dB & 12 dB & 16 dB \\
\hline
\textbf{MLD} 
& 0.054378 & 0.015725 & 0.003014 & 0.000463 & 0.000113
& 0.022834 & 0.005125 & 0.001149 & 0.000181 & 0.000083 \\
\textbf{MLP}        
& 0.055550 & 0.014987 & 0.003024 & 0.000699 & 0.000122
& 0.022565 & 0.004641 & 0.000972 & 0.000167 & 0.000022 \\
\textbf{CNN}        
& 0.055620 & 0.014990 & 0.003031 & 0.000707 & 0.000122
& 0.022621 & 0.004662 & 0.000974 & 0.000172 & 0.000023 \\
\textbf{ResNet}     
& 0.055761 & 0.015022 & 0.003021 & 0.000702 & 0.000121
& 0.022656 & 0.004676 & 0.000977 & 0.000167 & 0.000021 \\
\hline
\end{tabular}
\end{table*}

When the SISO results for $m=1$ are examined, all methods show nearly identical performance across all SNR values. At 16 dB, the BER values range from 0.009418 (CNN) to 0.009554 (MLD), with negligible differences among methods. The highest BER is obtained by the MLD method, while CNN shows marginally better performance. For $m=2$, similar behavior is observed, with all methods achieving comparable BER performance around $4.9 \times 10^{-3}$ at 16 dB SNR. In the SISO configuration, the absence of spatial diversity results in nearly equivalent detection performance across all methods.

For the MIMO configuration ($N_T=2$, $N_R=2$), significant performance improvement is observed due to spatial diversity gains. Under $m=1$ fading at 16 dB, MLD achieves the lowest BER of $1.13 \times 10^{-4}$, followed closely by the DL-based methods with values around $1.21 \times 10^{-4}$ to $1.22 \times 10^{-4}$. For $m=2$, MLD achieves $8.3 \times 10^{-5}$, while DL-based methods achieve approximately $2.1 \times 10^{-5}$ to $2.3 \times 10^{-5}$. Overall, the results demonstrate that both detector types ensure reliable performance in the considered MIMO setting, while the DL-based methods achieve this performance with markedly reduced computational complexity.

Additionally, to evaluate the impact of modulation order on detection performance, the proposed MLP-based detector is tested with both 4-QAM and 16-QAM modulation schemes under $m=2$ fading conditions. As shown in Fig.~\ref{fig:16qam}, the MLP-based detector maintains near-optimal performance for both modulation orders. For 4-QAM, the BER curves of MLD and MLP-based detectors are nearly indistinguishable across all SNR values. Similarly, for 16-QAM, the MLP-based detector achieves performance comparable to the MLD benchmark, confirming the robustness of the proposed approach across different constellation sizes.

\begin{figure}
    \centering
    \begin{minipage}{\columnwidth}
        \centering
        \includegraphics[width=.9\linewidth]{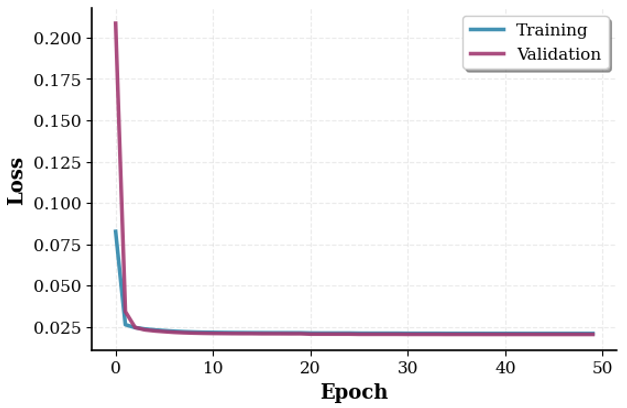}\\
        (a) Training and validation loss convergence.
    \end{minipage}
    
    \vspace{0.3cm}
    
    \begin{minipage}{\columnwidth}
        \centering
        \includegraphics[width=.9\linewidth]{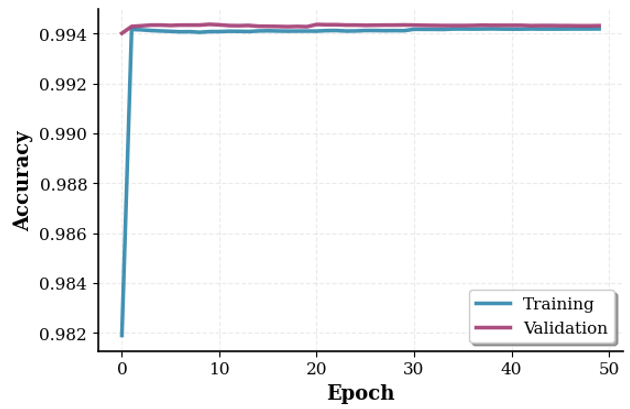}\\
        (b) Training and validation accuracy progression.
    \end{minipage}
    \caption{Training and validation loss and accuracy curves of the MLP model for one fold.}
    \label{fig:mlp_acc_loss}
\end{figure}

These evaluations show that the DL-based detectors used in this study achieve near-optimal BER performance while maintaining significantly lower computational complexity compared to MLD. Additionally, in terms of complexity cost, MLP is substantially less expensive than the classical MLD method. Their lower complexity and comparable detection performance make these networks attractive for practical OTFS-SD tasks.

Among the DL-based methods, MLP and CNN showed the best results in the study for both $m=1$ and $m=2$. However, the ResNet network, which also belongs to the CNN family and employs skip connections to propagate feature maps across layers, did not yield any meaningful performance gain in our experiments. Although such architectures are generally effective at preserving gradient flow and mitigating vanishing effects, this advantage did not translate into improved BER performance under the considered setup and not provide a significant advantage over the simpler MLP and CNN architectures.

A possible explanation is that the characteristics of the dataset and task did not provide additional benefit for deeper representations, which may have limited the model’s ability to generalize beyond the performance of simpler architectures. Moreover, particularly for low $MN$ values, the high complexity of ResNet makes using such a deep network in this specific study unnecessary and costly. The MLP architecture, with its significantly lower computational complexity and near-identical performance, emerges as the most practical choice for OTFS signal detection in the investigated scenarios.
As shown in Fig.~\ref{fig:mlp_acc_loss}, both the training and validation loss curves rapidly converge, 
and the validation accuracy remains consistently high across epochs. 
This indicates that the MLP structure generalizes well to the task without exhibiting signs of overfitting. 
Therefore, the adopted architecture provides sufficient representational capacity, 
and employing deeper or convolution-based models would not yield additional benefits for this scenario.

\section{Conclusion and Future Work}
\label{sec:conclusion}
This study proposes a two-stage, low-complexity DL–based SD framework for MIMO-OTFS systems, where the signals collected at multiple receive antennas are first combined via MRC and then detected either by an MLD detector or by DNNs. Simulation results under Nakagami-$m$ fading show that the proposed MLP, CNN, and ResNet-based detectors achieve near-optimal BER performance, closely approaching the MLD benchmark in both SISO and MIMO scenarios. At the same time, complexity analysis reveals a clear advantage for the MLP-based detector: it maintains linear complexity $\mathcal{O}(MN)$ and significantly reduces the computational complexity compared to CNN, ResNet, and MLD, while preserving robust detection accuracy. These results indicate that relatively simple DL architectures, when combined with MRC, are sufficient for effective OTFS-SD, making the MLP-based design a highly practical and efficient solution for latency and energy-constrained 6G wireless applications. 

Although this study shows that DL-based detection makes the symbol decision stage computationally simpler, the preprocessing operations, especially the effective channel matrix computation and MRC preprocessing, which require $\mathcal{O}(N_T N_R (MN)^3)$ operations, still dominate the overall receiver complexity. Future research should focus on creating end-to-end DL architectures that can reduce the computationally demanding matrix operations. Additionally, the proposed DL-based architecture can be expanded to solve the computational issues given by upcoming OTFS-based systems. For example, reconfigurable intelligent surface (RIS)-assisted OTFS systems add complexity through phase shift optimization across reflecting elements, whereas integrated sensing and communication (ISAC)-OTFS systems necessitate cooperative processing of communication and sensing inputs. DL methods can give unified solutions to these problems by learning optimal detection and resource allocation strategies without the use of explicit iterative algorithms. Furthermore, advanced machine learning paradigms such as federated learning allow for privacy-preserving distributed training across heterogeneous devices in massive Internet of Things (IoT) deployments, while reinforcement learning can adaptively optimize detection parameters in dynamic high-mobility environments. Extending the proposed methods to massive MIMO-OTFS configurations with hundreds of antennas, along with investigating model-driven hybrid architectures that combine signal processing insights with data-driven learning, will be essential for realizing computationally efficient receivers for diverse 6G wireless applications.

\bibliographystyle{IEEEtran}
\bibliography{Referanslar}
\end{document}